\documentclass[a4paper,fleqn,usenatbib,useAMS]{mnras}
\usepackage{color}
\usepackage{graphicx}	\usepackage{amsmath}	\usepackage{amssymb}	\usepackage{multicol}        \usepackage{bm}		\usepackage{pdflscape} \usepackage{arydshln}

\usepackage[T1]{fontenc}
\usepackage{ae,aecompl}

\usepackage{txfonts}

\title[Gravitational energy spectra]{Probing the multi-scale interplay between gravity and turbulence
 -- Power-law like gravitational
energy spectra of the
Orion Complex}
\author[Guang-Xing Li,  Andreas Burkert]{Guang-Xing Li$^{1}$\thanks{Contact e-mail:
\href{mailtogx}{gxli@usm.lmu.de}} ,  Andreas
Burkert$^{1, 2}$
\\
$^{1}$University Observatory Munich, Scheinerstrasse 1, D-81679 M\"unchen,
Germany\\
$^{2}$ Max-Planck-Fellow, Max-Planck-Institute for Extraterrestrial
Physics, Giessenbachstrasse 1, 85758 Garching, Germany
}
\date{\today}

\pubyear{2015}

\begin{document}
\label{firstpage}
\pagerange{\pageref{firstpage}--\pageref{lastpage}}
\maketitle

\begin{abstract}
Gravity plays a determining role in the evolution of the molecular
ISM.
In \citet{2016arXiv160304342L}, we proposed a measure called gravitational energy
spectrum to quantify the importance of gravity on multiple physical scales.
 In this
work, using a wavelet-based decomposition technique, we
derive the gravitational energy spectra of the Orion A and the Orion B
molecular cloud from observational data.
 The gravitational energy spectra exhibit power-law-like
 behaviours.
From a few pc down to $\sim  0.1 $ pc scale, the Orion A and Orion B molecular
cloud have $E_{\rm p}(k)\sim k^{-1.88}$ and $E_{\rm
 p}(k)\sim k^{-2.09}$, respectively. These scaling exponents are close to the
 scaling exponents of the kinetic energy power spectrum of compressible
 turbulence (where $E\sim k^{-2}$), with a near-equipartition of turbulent
 versus gravitational energy on multiple scales.
 This provides a clear evidence that gravity is able to counteract effectively
 against turbulent motion for these length scales. The results confirm our
 earlier analytical estimates. For the Orion A molecular cloud, gravity inevitably dominates over turbulence inside the cloud. 
Our results provide a clear observational proof that gravity is playing a
determining role in the evolution these molecular clouds from the
cloud scale down to $\sim 0.1\;\rm pc$. { However, turbulence is
likely to dominate in clouds like California.  } The method is general and
should be applicable to all the astrophysical problems where gravity plays a role.
\end{abstract}


\begingroup
\let\clearpage\relax
\endgroup
\newpage

\section{Introduction}
Star formation takes place in the dense and shielded parts of the 
interstellar medium \citep[see e.g. ][for a review]{2014prpl.conf....3D}.
Magnetic fields are also believed to be playing a role
\citep{2014prpl.conf..101L}. Gravity is a long-range force, and plays a
determining role in molecular cloud evolution. Observational constraints on gravity in
star-forming regions are thus crucial for understanding the star formation
process. However, it is not even clear if the dynamics on
the cloud scale is dominated by gravity or not \citep{2009ApJ...699.1092H,2011arXiv1111.2827K}.
Part of the difficulty comes from the existence of complicated structures in the
molecular interstellar medium (ISM)
\citep{1979ApJS...41...87S,2000prpl.conf...97W,2008ApJ...680..428G,2010A&A...518L.103M}.

The virial parameter \citep{1992ApJ...395..140B} is a measure of
the intensity of  gravity in terms of its energy  measured with respect to
energy from e.g.
turbulent motion. To evaluate a virial parameter, one needs to define an
object for which it is calculated, and { is dependent on other
implicit assumptions on the underlying dynamics \citep{2006MNRAS.372..443B}.} This
makes it difficult to provide constraints of gravity over multiple scales. A few attempts have been made to overcome the limitation of
the virial parameter. One can evaluate the virial parameter on different physical scales by applying it to a dendrogram representation of the observational data \citep{2009Natur.457...63G,2008ApJ...679.1338R}. Recently, \citet{2015A&A...578A..97L} proposed a method called G-virial which generalises the concept of the gravitational potential to observations in the
position-position-velocity (PPV) space. This allows one to analyse
the structures of molecular gas with the G-virial maps derived from the
observations. They found that molecular clouds are close to gravitationally
bound when the boundaries of the regions are determined from the G-virial map.
These approaches provide constraints on the importance of gravity compared to
the observed turbulent motion over multiple scales.
  A complimentary approach is to study the \emph{effects} of
gravity. The acceleration mapping method \citep{2016arXiv160305720L} is
developed along this line of thought. The method computes gravitational
acceleration based on maps of surface densities. It is found that
gravitational acceleration behaves in a non-uniform fashion, and this can lead
to fast collapse in localised regions in molecular clouds
\citep{2004ApJ...616..288B, 2016arXiv160304342L}.  

  What remains unconstrained is the distribution of gravitational
energy across scales. This aspect is
important for several theoretical reasons: first, how energy distributes across
scales is a fundamental measure of a system ( e.g. one fundamental way to
describe the electromagnetic wave is to construct is spectrum). A second
motivation is that supersonic turbulence is believed to
be important in the dynamics of molecular clouds \citep{2004RvMP...76..125M}.  The
distribution of kinetic energy of the turbulent motion can be derived
theoretically, and it obeys roughly $E \sim k^{-2}$ where $k$ is the wavenumber.
One can study the interplay between turbulence and gravity over multiple scales, provided that a similar measure of gravitational energy can be
obtained. 


Along this line of thought, \citet{2016arXiv160304342L} proposed a formalism to
represent the multi-scaled structure of the  interstellar medium
(ISM) gravitational field and to quantify its impact on cloud evolution by
constructing a quantity called \emph{gravitational energy spectrum}. The derived gravitational energy spectra are directly comparable to
e.g. the turbulence power spectra. By studying the gravitational energy spectra
of a sample of 8 molecular clouds, \citet{2016arXiv160304342L} found that
molecular clouds are close to a state where the kinetic energy of turbulence and
gravitational energy reaches a near equi-partition.
For star-forming clouds, gravity gradually takes over turbulence as one moves from larger to smaller scales. 

The fact that the gravitational energy can balance and even dominate over
turbulent motion in star-forming clouds has important implications in cloud evolution
theory.
It provides support for cloud evolution models such as the
hierarchical gravitational collapse model
\citep{1953ApJ...118..513H,1993ApJ...419L..29E,2003ApJ...585L.131V}. See also
\citet{2013ApJ...773...48B}. These theoretical possibilities should be further
explored.
However, in \citet{2016arXiv160304342L}, the results are inferred from the
surface density PDF, where a shell model was assumed. Although the authors provided justifications
why these assumptions are valid, a more detailed study is desired to validate
the simplications made in their model.

We provide observational constraints on the
importance of gravity by combining: (a) state-of-the-art observations of the
surface density structure of molecular clouds constructed from ESA's Planck and
Herschel observations \citep{2014A&A...566A..45L}, (b) constructing 3D volume density
distributions from observational data, the method we use is an improved
version of the method described in \citet*{2014Sci...344..183K}, (c) a formalism
to quantify the importance of gravity on a multiple of scales, proposed
 in \citet{2016arXiv160304342L} and (d)  a wavelet-based technique to decompose
 the gravitational field into multiple components.

We study the gravitational energy spectra of the clouds in the Orion molecular
cloud complex.
The cloud is the most massive active star-forming complex in the solar neighbourhood
\citep{2008hsf1.book..459B}, and is subject to intensive studies. 
Based on the slope of the surface density PDFs derived in
\citet{2014A&A...566A..45L}, \citet{2016arXiv160304342L} obtained constraints on the
gravitational energy spectra for both the Orion A and the Orion B molecular
cloud. They conclude that Orion A molecular cloud is in a stage where gravity
takes over turbulence on smaller scales, and for Orion B they found that the
cloud is close to a critical state where the gravitational energy and
turbulence kinetic energy reaches equi-partition.

In this paper, we propose a wavelet-based method to construct multi-scale
gravitational energy spectra from observational data. The method is general, and
can be applied to structures with arbitrary geometries.  In Sec.
\ref{sec:concept} we introduce the concept of the gravitational energy spectrum.
We
reconstruct the 3D density structure of the Orion molecular clouds from the
observational data (Sec.
\ref{sec:observatioins}).
Gravitational potentials of the clouds are constructed based on the 3D density
structures  (Sec. \ref{sec:phi}). 
A wavelet-based decomposition technique is used
to obtain constraints on gravity over a multitude of physical scales
(Sec. \ref{sec:wavelets}). The results are presented in Sec.
\ref{sec:results}.
In Sec.
\ref{sec:test} we compare the observed gravitational energy spectra with the
gravitational energy spectra of a Mach 5.6 compressible turbulence. This also
allows us to access the performance of our method under a situation where the
underlying geometry is sufficiently complicated. In Sec. \ref{sec:comp} we
compare our method with other pre-existing methods. 
 In Sec.
\ref{sec:conclu} we conclude.

\section{Probing cloud fragmentation with gravitational energy spectra}
\label{sec:concept} 
\begin{figure*}
\includegraphics[width=\textwidth]{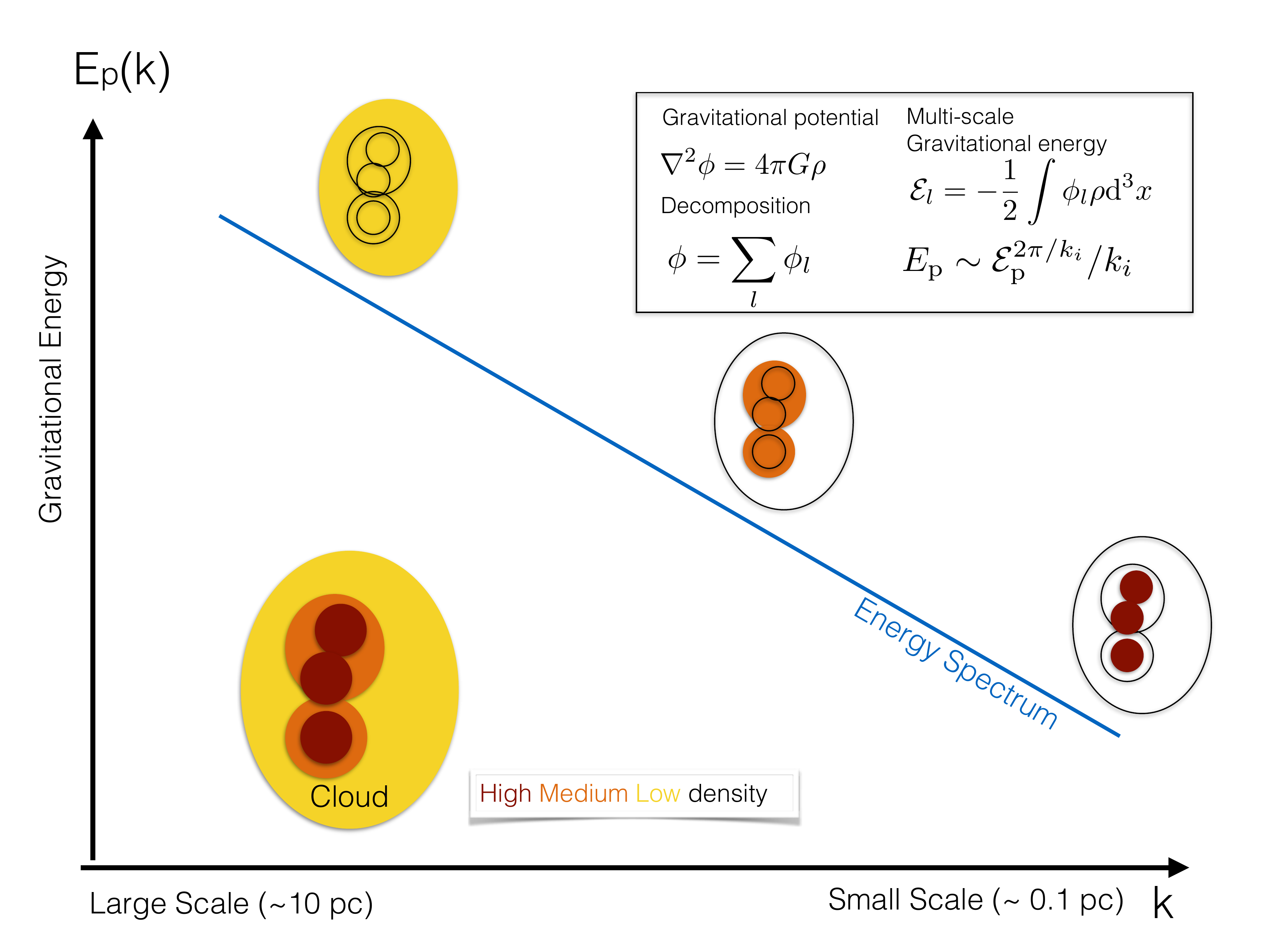}
\caption{\label{fig:concept} An illustration of the concept of the gravitational
energy spectrum. {  In an idealised picture, one can view an object
as being composed of a set of nested structures.}
Smaller structures (in red) tend to have higher densities, and tend to reside in
larger structures (in orange and yellow). In our gravitational energy spectra
plot, the $x$-axis is the wave number $k=2\pi/l$, which is inversely proportional to the
length scale $l$. The $y$-axis is the  gravitational energy
density (which has a unit of energy divided by the wave number $k$).
Gravitational energy on the large scale is usually contributed from structures
of larger sizes (in yellow), and gravitational energy on small scales is
contributed from structures of small sizes (in red). See  Sec.
\ref{sec:concept} for explanations. The detailed mathematical formula for
evaluating the gravitational energy spectra are listed in the box. The formulas
are presented and explained in Sec.
\ref{sec:wavelets}. }
\end{figure*}

The meaning of the gravitational energy spectrum is illustrated in Fig.
\ref{fig:concept}. In astrophysical settings, a density distribution $\rho(x, y,
z)$ is composed of structures of different sizes. 
Smaller structures tend to have higher densities, and tend to reside inside
larger structures. On the coarsest (largest) scale, one can measure
the masses $m_0$ and sizes $l_0$ (which satisfy $l_0^{\rm min}<l_0<l_0^{\rm
max}$ where $l_0^{\rm min}$ and $l_0^{\rm max}$ are the lower and upper bounds
of the scale $l_0$) of the structures, and the
gravitational energy of the cloud on the coarsest scale is $\mathcal{E}_{l_0} =
G m_{l_0}^2 / l_0$. On a smaller scale, the cloud is composed of sub-regions of
sizes $l_1^{\rm min}<l_1<l_1^{\rm max}$ and masses $m_{l_1(i)}$, and the total gravitational energy on this scale is
$\mathcal{E}_{l_1} =\sum_i G m_{l_1(i)}^2 / l_1$. One can further increase the
resolution, and compute the gravitational energy on even smaller scales e.g. $l_2, \ldots\, l_n$ until the
resolution limit of the data has been reached. By investigating the dependence
of $E_l$ on $l$ one can obtain a multi-scale picture of gravitation energy in
molecular clouds.  For a cloud that has a simple morphology, this can be
achieved analytically \citep{2016arXiv160305720L}.
 Note that the picture here is
proposed to help achieving an intuitive understanding of the gravitational
energy spectrum. An equivalent yet mathematically rigours way to define the
gravitationally energy spectrum can be achieved by decomposing the gravitational
potential $\phi$ into components of different scales $\phi_l$ and evaluate $\int
\rho \phi_l {\rm d}^3 x$.
This is described in Sec. \ref{sec:wavelets}.

To achieve mathematical consistency, one also needs to normalise the
gravitational energy $\mathcal{E}$ by the range of physical scales within which
$\mathcal{E}$ is calculated. Either one has to work with $E_l =
\mathcal{E} / (l^{\rm max} - l^{\rm min})$ or with $E_k = \mathcal{E} /
(k^{\rm max} - k^{\rm min})$ where $k = 2 \pi / l$, $l$ is the scale and $k$ is
the wavenumber.  The gravitational energy spectra presented in this paper are
normalized according to $k$. Therefore $E_k $ has a unit of energy divided by 
wave number \footnote{Depending on the situation, one can also choose to further
normalise it by the mass $m$.}.

The gravitational energy spectrum
defined in this work is a generalisation of those discussed in \citet{2016arXiv160304342L}. The
improvement is the regions are now allowed to have arbitrarily complicated
geometries, and in \citet{2016arXiv160304342L} only close-to-spherical geometries are considered. 
 
\section{Observations and surface density modelling}\label{sec:observatioins}
We obtain observational data from \citet{2014A&A...566A..45L}. The map traces
column densities $ 1 \times 10^{21} {\rm cm^{-2}} < N < 2 \times 10^{23} {\rm
cm^{-2}}$, and resolve down to $\sim 36 \arcsec$ (correspond to
0.07 pc at a distance of 414 pc \citep{2007A&A...474..515M}). 

\citet*{2014Sci...344..183K} developed a technique to construct 3D density
distributions from observational data. The idea is based on a simple assumption
that a smaller gas clump observed in 2D is related to a smaller clump in real
3D space.
Thus, the density structure of a molecular cloud can be reconstructed by
decomposing the observational map into a set of 2D ellipses and by linking the ellipses to 3D
ellipsoids.

\citet*{2014Sci...344..183K}  took a two-step approach to reconstruct 3D
 density distributions.
First, a 2D projected observational map is decomposed into components of
different sizes, generating a so-called size map. Second, each size map is
 decomposed into a set of structures with masses $M_i$ and areas
$A_i$.
The extent $H_i$ of each of the structure in the third dimension is estimated.
An ellipsoid in 3D can be reconstructed based on $M_i$, $A_i$ and $H_i$ for each structure. 

In this work, we reconstruct 3D density distributions using a method that
shares a similar spirit with that of \citet*{2014Sci...344..183K}. However, several
critical improvements have been made. First, we have developed an improved
method to decompose the map into contributions from structures of different
sizes, generating the ``size map''.
\citet*{2014Sci...344..183K}  used a wavelet-based method to construct the size
 map. However, for the Herschel-Planck surface density map used in this work,
 because of the improved dynamical range,  the
 wavelet-based method produces negative artefacts around regions where the
 column densities have been significantly enhanced. We thus developed an
 improved decomposition method where significant structures of different scales
 are identified by an iterative approach.
 The size map derived from our improved method is completely free of negative
 artefacts. Details of the improved method can be found in Appendix \ref{sec:size:map}.

\begin{figure}
\includegraphics[width = 0.42 \textwidth]{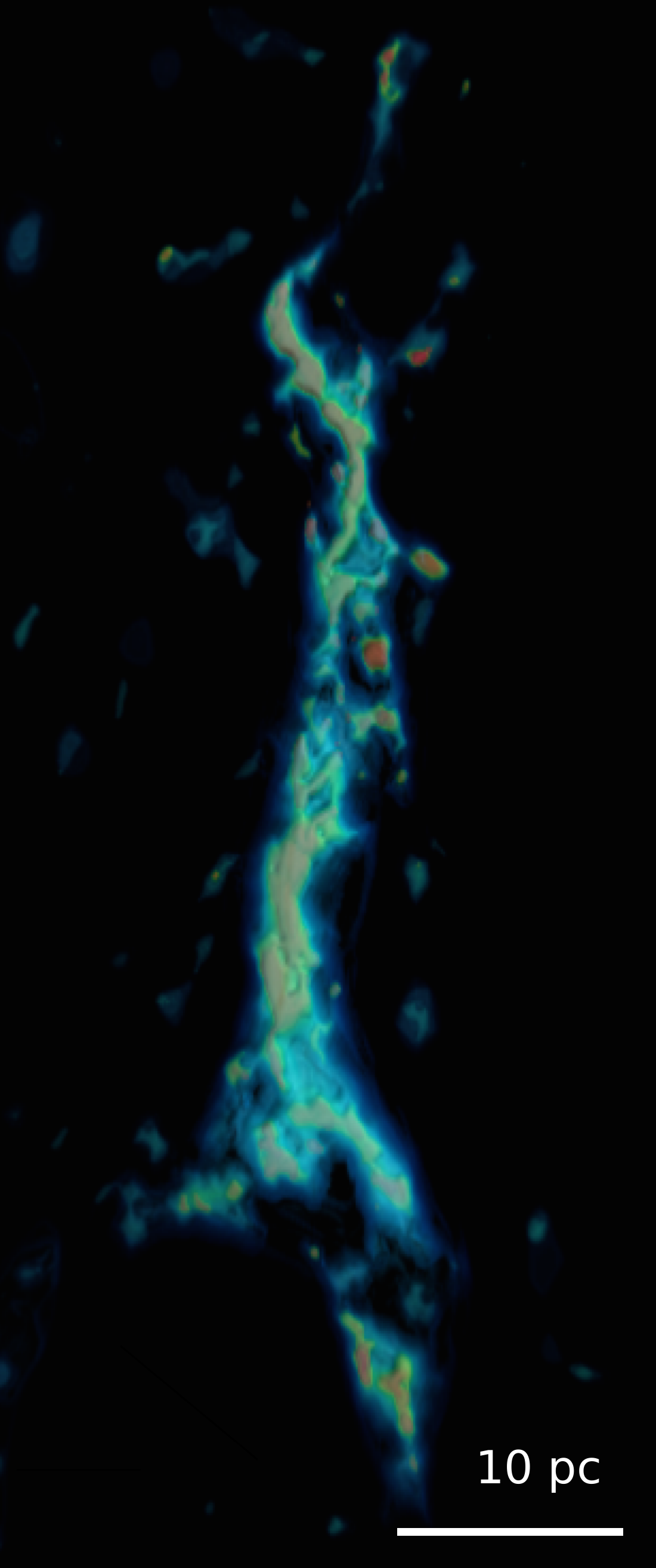}
\caption{A 3D volume rendering of the reconstructed volume density structure of
the Orion A molecular cloud. The rendering is sensitive to gas with $n_{\rm
H_2} \sim 2 \times 10^4\;\rm cm^{-3}$. Filamentary structures
 are preserved in our new reconstruction. {A scalebar is included in
 the plot.}
\label{fig:render}}
\end{figure}

 In the second step, structures are identified from the size map. Properties
 such as masses and sizes are evaluated. 
 This information is then used to construct
 a 3D volume density map. \citet*{2014Sci...344..183K} used the
 algorithm ``clumpfind'' \citep{1994ApJ...428..693W} to detect structures.
Masses and sizes of these structures are estimated. By assuming
 that they are 3D ellipsoids, a 3D volume density map can be reconstructed. In
 the formalism of \citet*{2014Sci...344..183K}, the ellipsoids have sharp
 boundaries. These artificial boundaries do not affect the volume density PDF, but are not
 desirable for other applications. 
 To overcome this difficulty, we have implemented a contour-based algorithm
 called contour-decompose to reduce the artefacts produced at the boundaries of
 the ellipsoids.
 The details of the algorithm can be found in Appendix \ref{sec:3d}. The improved algorithm produces new
 data sets that are better for general-purpose analysis.
 In Fig.
 \ref{fig:render} we present a volume rendering of the reconstructed 3D volume
 density map of the Orion A molecular cloud. { Filamentary structures
 are preserved in our new reconstruction.}

\section{Computation of the gravitational field}\label{sec:phi}
 The
gravitational potential is computed by solving the Poisson Equation $\nabla^2
\phi = 4 \pi G \rho$. It can be solved efficiently in the Fourier space
according to 
\begin{equation}
\phi_k = \frac{- 4 \pi G}{k^2} \rho_k \;,
\end{equation}
where $\phi_k$ and $\rho_k$ are the Fourier transforms of $\phi(x, y, z)$ and
$\rho(x, y, z)$ respectively.

Computing gravitational potentials in the Fourier space automatically assumes
that the boundary conditions are periodic. To minimise the effects of this
limitation to our results, zero density paddings are added to the 3D volume
density map before the computation.
Note also that molecular clouds are never isolated but part of an
inter-connected network of molecular structures that might affect the clouds
gravitationally. {  This would have an effect on the gravitational
energy spectrum at scales larger than e.g. half the box size.}
The volume density map after the padding procedure has a size of $(N, N, N)$ where $N$ is the maximum of the lengths of the 3 axes of the
input data.

\section{Deriving the gravitational energy spectra}\label{sec:wavelets}
We evaluate the gravitational energy spectra from observational data based on
the reconstructed 3D volume density map $\rho(x,y, z)$ and the gravitational
potential $\phi(x, y, z)$.

 We employ a two-step approach. First, 
 we use the wavelet transform \emph{a trous} \citep[see e.g.
 ][]{1998ipda.book.....S} to decompose the gravitational potential $\phi$ into
 components of different physical scales $\phi_l$ (where $l$ denotes the
 characteristic length scale of a particular wavelet component):
\begin{equation}\label{eq:wavelet}
\phi(x, y, z) = \sum_{l} \phi_{l}(x, y, z) + R(x, y, z)\;,
\end{equation}
where $l$ represents the physical scale. $l_i = 2^i \, {\rm d}x$ where $i=0, 1,
2, \ldots, n $, and ${\rm d}x$ is the map resolution. $R(x, y, z)$ is the residual.
\footnote{ We use the Mexican hat
wavelet, \url{https://en.wikipedia.org/wiki/Mexican_hat_wavelet}.}.
 { 
The different components are obtained by taking the Fourier modes at
different intervals of spatial frequencies, and the wavelet component at scale
$l$ is contributed from Fourier modes whose scales are limited between
$l/\sqrt{2}$ and $\sqrt{2}\; l$. Because long, filamentary structures have plenty of fluctuations on scales comparable to their
widths, in the formalism of the wavelet transform, they can appear on scales
that are comparable to their widths.}

For each 3D gravitational potential map of size $(N_{\rm x}, N_{\rm y}, N_{\rm
z})$, we added zero paddings around the map so that a new map is
created with a size $(N, N, N)$ where $N = {\rm max}(N_{\rm x}, N_{\rm y}, N_{\rm
z})$
is created. The minimum physical scale is limited by the map resolution.
Because we do not have information of structures of
gas beyond the observed map, the maximum physical scale is determined by
$l= 2^n {\rm d}x < {\rm min}(N_{\rm x}, N_{\rm y}, N_{\rm z}) / 2 \times {\rm
d}x$ .

Since $l_i = 2^i {\rm
d}x$ where $i=0, 1, 2, \ldots, n $, for a given $l_i $, $E_{\rm
p}^{l}$ represents the gravitational energy distributed between
$l_{\rm min}= l_i/\sqrt{2}$ and $l_{\rm max} = l_i \times \sqrt{2}$.

The gravitational binding energy $\mathcal{E}_{\rm i}^{l}$  of the wavelet
component of length $l$ is
\begin{equation}\label{eq:int}
\mathcal{E}_{\rm p}^{l} = - \frac{1}{2} \int   \rho\; \phi_l\; {\rm
d}x {\rm d}y {\rm d}z \;,
\end{equation}
which ensures that $E_{\rm p}^{\rm tot} = \sum_l \mathcal{E}_{\rm p}^l$ where
$E_{\rm p}^{\rm tot}$ is the total gravitational binding energy of the cloud.

The gravitational energy per given length is defined as
\begin{equation}\label{eq:el}
E_{\rm p}^{l \rightarrow l_i} = \frac{\mathcal{E}_{\rm p}^{l}}{l_{\rm max} -
l_{\rm min}} = \frac{\mathcal{E}_{\rm p}^{l_i}}{(\sqrt{2}-\sqrt{2}^{-1})\times
l_i} \approx \frac{\mathcal{E}_{\rm p}^{l_i}}{0.7 \times l_i}
\end{equation}

It is more convenient to convert Eq. \ref{eq:el}  into $E_{\rm
p}(k)$ where $k$ is the wavenumber ($k_i = 2 \pi /l_i$):
\begin{eqnarray}
E_{\rm p}^{k  \rightarrow k_i} &=&  E_{\rm p}^{l \rightarrow 2 \pi / k_i} \;
\frac{2 \pi}{k_i^2}\\
\nonumber &\approx& 1.4 \times
\frac{\mathcal{E}_{\rm
p}^{l \rightarrow 2 \pi / k_i} } { k_i}\;,
\end{eqnarray}
where we have used $|{\rm d} l / {\rm d} k| = 2 \pi / k^2$, and $E_{\rm p}^l$
was defined in Eq. \ref{eq:el}.
Following \citet{2016arXiv160304342L}, we call $E_{\rm p}(k)$ the \emph{gravitational
energy spectrum} of a cloud.

$E_{\rm p}(k)$ contains information on how the gravitational energy of a
molecular cloud is distributed among different physical scales. Its unit is the
same as that of  the turbulence power spectrum. By comparing $E_{\rm p}(k)$ with
the turbulence power spectrum one can evaluate the relative importance of turbulence and gravity on different physical scales.

\section{Results}\label{sec:results}
\citet{2014A&A...566A..45L} found that the Orion A and Orion B molecular clouds
have different scaling exponents of the surface density PDFs, implying that they are
probably at different evolutionary stages. In this work, we treat them as
different objects, and study their gravitational energy spectra separately.

\begin{figure*}
\includegraphics[width = 0.95 \textwidth]{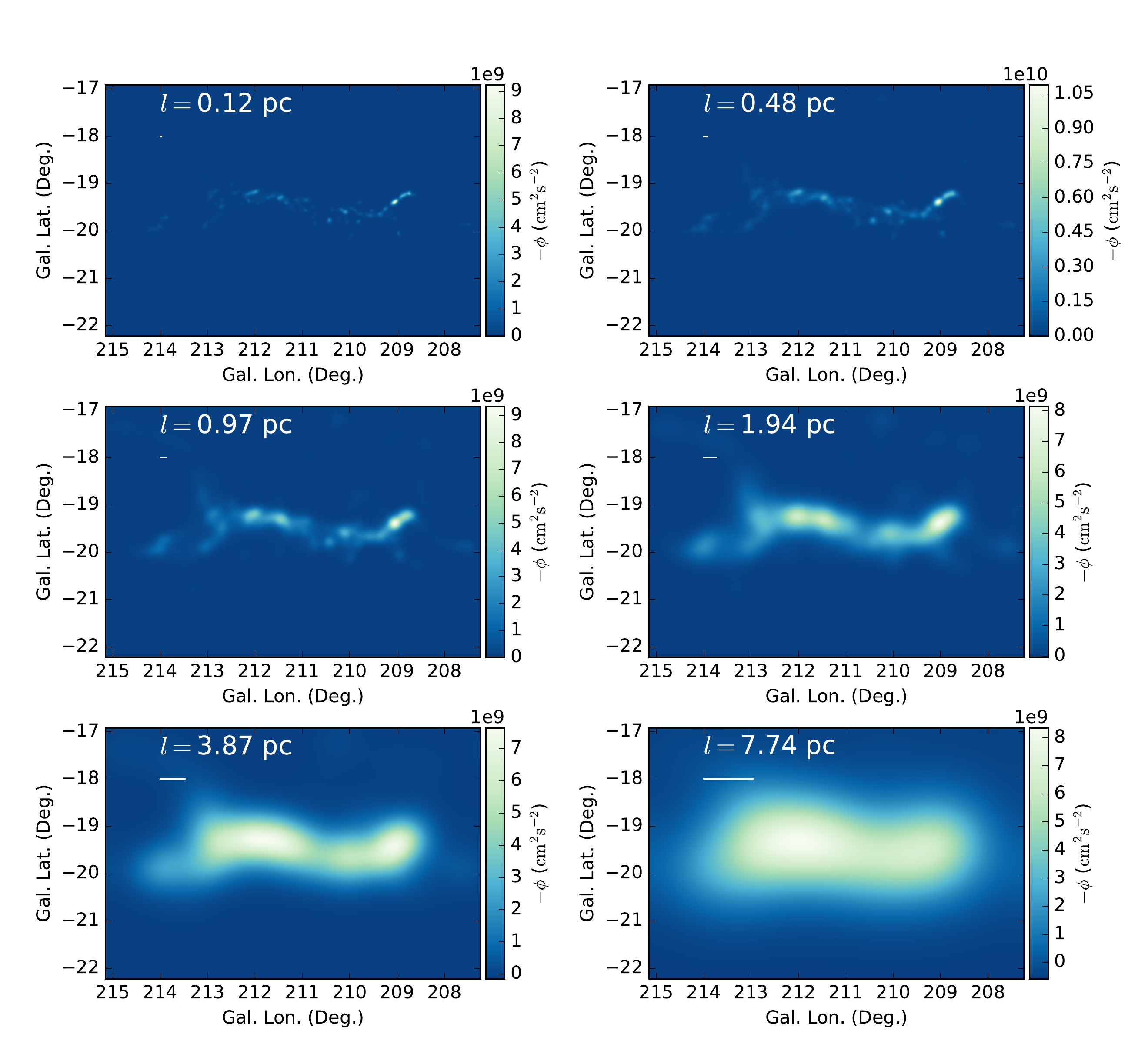}
\caption{\label{fig:componets}Visualisation of several wavelet components
$\phi_l$ of the gravitational potential of the Orion A molecular cloud, { where
$l$ is defined in Eq. \ref{eq:wavelet}}, { i.e. the sequence $l =
0.12\;\rm pc$, $l = 0.48\;\rm pc$, \ldots, $l = 7.74\;\rm pc$ correspond to the rough scale of
the corresponding wavelet components. Scalebars that indicate the relevant
scales are added to the panels. } The wavelet components of the
gravitational potential are defined in 3D space. The images shown here are
produced by taking the minimum of the gravitational potential $\phi(x,y,z)$ along each line of sight.
In each panel, $l$ indicate the relevant physical scale of the
corresponding wavelet component.  When $l$ is  small the component is
highly structured, and when $l$ is large the component is smoother.
{Here, one degree correspond to 7.5 pc.} The size of the region is
$\sim 50$ pc.
}
\end{figure*}

The gravitational binding energy of a wavelet component of length $l$ is defined
in Eq. \ref{eq:wavelet}, and a visualisation of $\phi_{ l}(x, y, z)$  provides
information concerning the spatial distribution of the gravitational binding energy that
belongs to component $l$.
In Fig. \ref{fig:componets} we present visualisations of a few wavelet components
of the gravitational potential of the Orion A molecular cloud. When $l$ is
small, the gravitational binding energy comes from highly filamentary, thin gas
structures in the cloud; and when $l$ is large, the gravitational binding energy
is contributed from a smoother distribution.

\begin{figure*}
\includegraphics[width = 0.95 \textwidth]{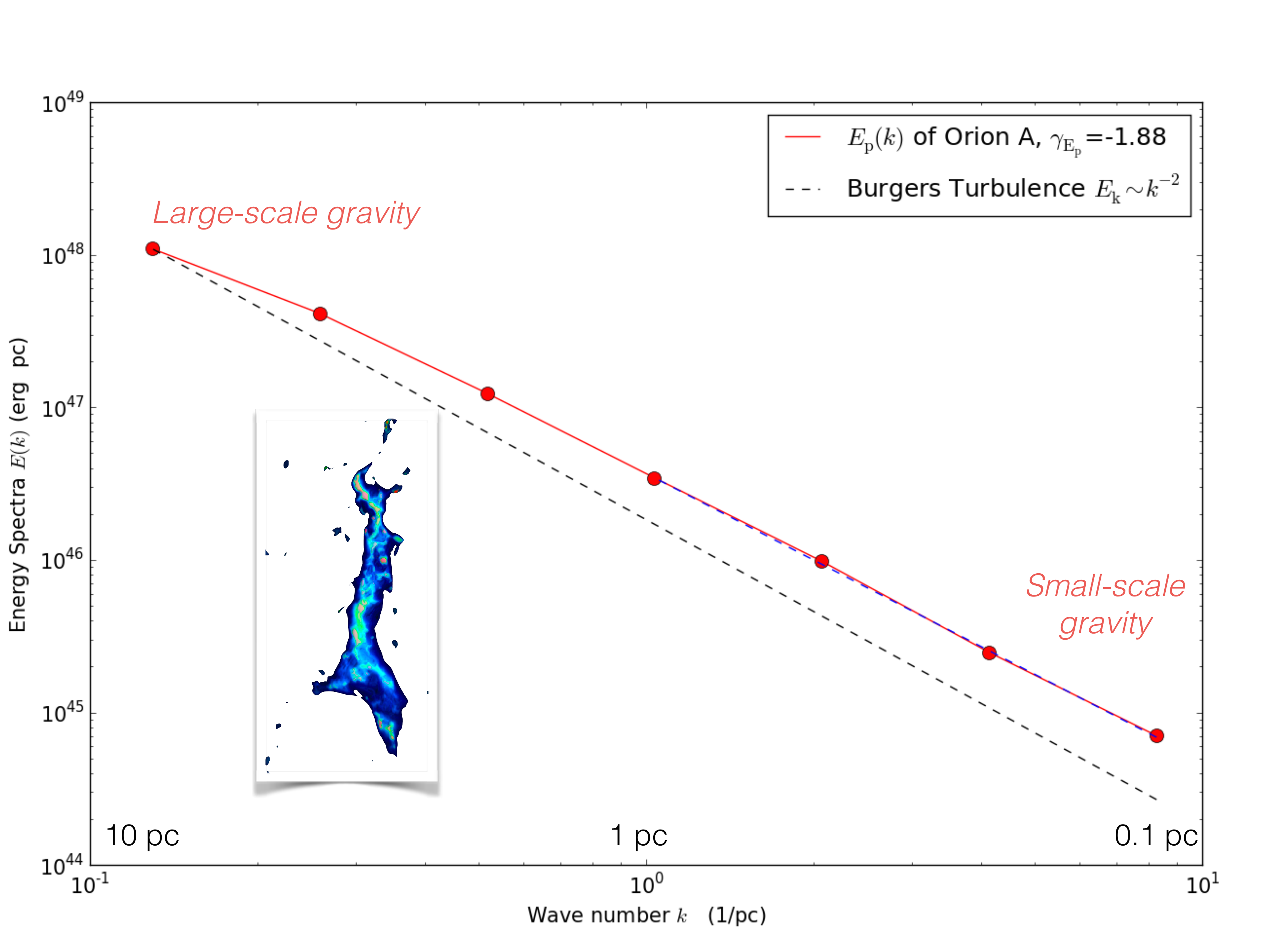}
\caption{\label{fig:orionia} Gravitational energy spectrum of the Orion A
molecular cloud. The energy spectrum is shown by the red line. The red
cycles show directly the measurements.  The $x$-axis is the wave number $k =
2\pi/l$, $l$ is the length scale, and the $y$-axis is the energy spectra.
$E_{\rm p}(k)\sim k^{-1.88}$ (dashed curve, superimposed on the red curve)
provides a fairly good approximation of the gravitational energy spectrum below the parsec scale. 
This is indicated as the blue dashed line. 
The black
dashed line is the velocity power spectrum of Burgers turbulence. A
volume rending of the density structure of the Orion A molecular cloud is
included as an inset plot.\label{fig:oriona} }
\end{figure*}

\begin{figure*}
\includegraphics[width = 0.95 \textwidth]{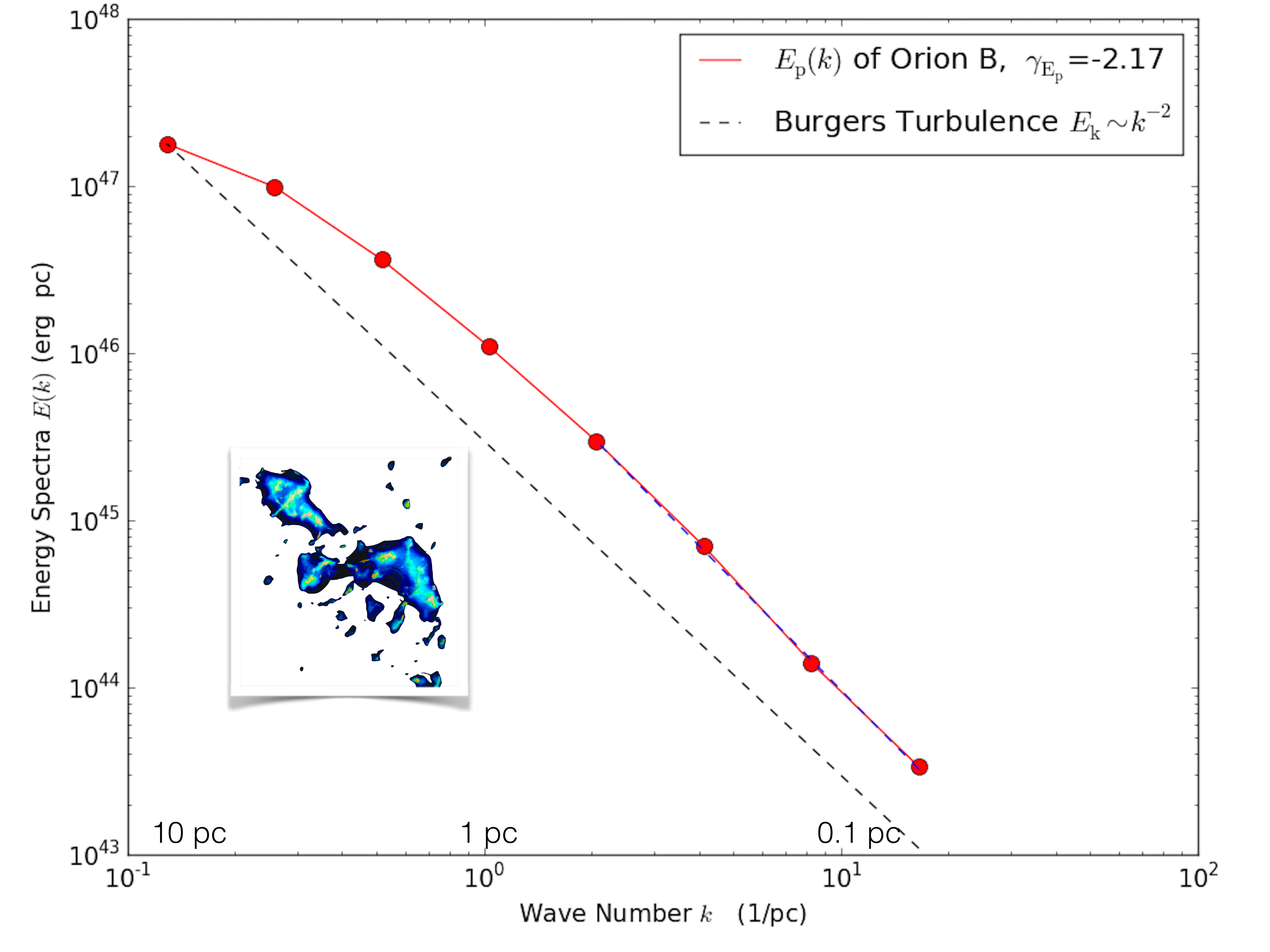}
\caption{\label{fig:orionia} Gravitational energy spectrum of the Orion B
molecular cloud. The energy spectrum is shown by the red line. The red
cycles show directly the measurements.  The $x$-axis is the wave number $k =
2\pi/l$, $l$ is the length scale, and the $y$-axis is the energy spectra.
$E_{\rm p}(k)\sim k^{-2.17}$ (dashed curve, superimposed on the red curve)
provides a fairly good approximation of the gravitational energy spectrum below the parsec scale. 
This is indicated as the blue dashed line. 
The black
dashed line is the velocity power spectrum of Burgers turbulence. A
volume rending of the density structure of the Orion B molecular cloud is
included as an inset plot.\label{fig:orionb}}
\end{figure*}

The gravitational energy spectra of the Orion A and Orion B molecular clouds
are presented in Figs. \ref{fig:oriona}, \ref{fig:orionb}, respectively. For
both clouds, the gravitational energy spectra exhibit power law forms. The
scaling exponents are close to 2. We fitted power laws to these gravitational
energy spectra by taking all the measurements with $l \lesssim 1\rm pc$. 
For Orion A we find $E_{\rm p}(k)\sim
k^{-1.88}$ and for Orion B  $E_{\rm p}(k)\sim k^{-2.09}$.  A stepper slope means energy
decreases faster 
with decreasing length scales. Thus, on average, the Orion A
molecular cloud has more gravitational energy per unit wavenumber on
smaller scales as compared to Orion B.

On the largest scale,
molecular clouds are close to being gravitational bound
\citep{2010ApJ...723..492R, 2009ApJ...699.1092H}, and the amount of turbulent
energy is comparable to the gravitational binding energy.
Many of the star cluster-forming clumps in the clouds are also in apparent
virial equilibrium \citep{2012A&A...544A.146W}. 
\citet{2015A&A...578A..97L} have demonstrated that by carefully choosing the
boundaries of the regions and applying the standard virial analysis, the
cloud is much more gravitationally bound compared to the result from a direct
virial analysis for the while cloud. A direct connection between gravity and
observed turbulent motion can also be seen from the $\alpha_k-\alpha_G$
plot \citep{2015arXiv151103670T}.
Accretion has been considered as a primary source that drives
turbulence in the cloud \citep{2010A&A...520A..17K}. It this is the case, then one does not expect
clouds to  deviate from the virial equilibrium by much. Recent simulations
also indicate that gravitational collapse is enough to explain the observed
level of turbulent motion \citep{2015arXiv151105602I}. Thus, we assume that
gas is gravitational bound on the cloud scale.

The scaling exponent of the gravitational energy spectra can be
compared with the kinetic energy spectra of the Burgers
turbulence. For Kolmogorov turbulence  $E_{\rm turb}(k)\sim k^{-5/3}$ and for
Burgers turbulence $E_{\rm turb}(k)\sim k^{-2}$. The turbulence in molecular cloud is
believed to be closer to Burgers \citep*{2008ApJ...688L..79F}. Here, for both clouds, the
scaling exponents of the gravitational energy spectra are indeed close to the
turbulent kinematic energy spectra. {  To first order, this
implies a multi-scale equi-partition of turbulent and gravitational energy. }
 Thus, if on the large scale, turbulence and
gravity can reach rough equilibrium, then on any smaller scales gravitational
energy is
least comparable to the energy from the turbulent cascade. 
For the Orion A
molecular cloud, the gravitational energy 
spectrum is shallower than Burgers.
At sub-parsec scale the gravitational energy should 
should therefore become more important than turbulence, and
dominate the cloud evolution.

\citet{2016arXiv160304342L} gave an analytical expression, allowing one to infer the
gravitational energy spectrum from the observed surface density PDF of the
clouds.
Based on observations of \citet{2015A&A...576L...1L}, they derived the scaling
exponents of the gravitational energy spectra for both clouds. Here, we compare
the gravitational energy spectra derived using our wavelet-based analysis with
those derived using the observational-analytical approach in \citet{2016arXiv160304342L}.
The results are summarised in Table \ref{tbl:comp}. The scaling exponents
derived from observations are in excellent agreements with those derived from
the surface density PDFs using the analytical formula, to an accuracy of $\lesssim 10$ \%, suggesting that the formalism
developed in \citet{2016arXiv160304342L} does capture the essential features of the
cloud that are necessary for proper evaluations of $E_{\rm p}(k)$.

{ The gravitational energy spectrum is a flexibly measure. In
this work, we only demonstrated the computation of gravitational energy spectra
for clouds like Orion A and Orion B. In practise, it is also possible to evaluate
the gravitational energy spectra for individual sub-regions in the cloud.
This can be achieved by changing the integration range in Eq. \ref{eq:int} .
This would enable us to investigate the spatial variations of gravitational energy
spectra within a given cloud (similar the investigation of column density PDF evolution in
\citet{2015A&A...577L...6S}). We reserve this for a future work.  }

\section{Comparison with numerical simulations}\label{sec:test}
{ 
Given a density distribution, the gravitational energy spectrum is 
uniquely defined. The major uncertainties in our results come from the
reconstruction of the 3D density structure.

To access the uncertainty of such a reconstruction, we use simulations  
from \citet{2010A&A...512A..81F}. The simulations are carried out under pure
turbulent initial conditions, and no self-gravity is included. We have chosen
this simulation, because (a) turbulence seems to be important in molecular
clouds
\citep[e.g.][]{2007ApJ...665..416K,2011ApJ...730...40P,2005ApJ...630..250K,2015ApJ...807...67M},
and many numerical simulations of star formation are based on initial conditions that are generated from a turbulent box. It would be of theoretical interest to look into the gravitational energy spectrum from such a turbulent medium, and (b) the simulations seems to exhibit a similar level of complexity to what is actually observed. \footnote{ The simulation we have chosen is from a turbulent box without self-gravity.
However, as has been demonstrated in \citet{2013ApJ...777..173B},
gravity acts to modestly reduce the line-of-sight confusion. Thus, the
simulation is suitable for testing our method.}}

%

The simulations of \citet{2010A&A...512A..81F} are
available at 
\url{http://starformat.obspm.fr/starformat/project/TURB_BOX}. We used the
simulation with compressible forcing. The Mach number is $M=5.6$. The snapshot
was taken at $t = 5 T$.
The simulation is rescaled to a mean $\rm H_2$ density of $10^3$, a sound speed
of 0.5 $\rm km/s$, and a size of 10 pc. A clump of $\sim 1\;\rm pc$ in size was
separated from the simulation box.

\begin{figure}
\includegraphics[width = 0.5 \textwidth]{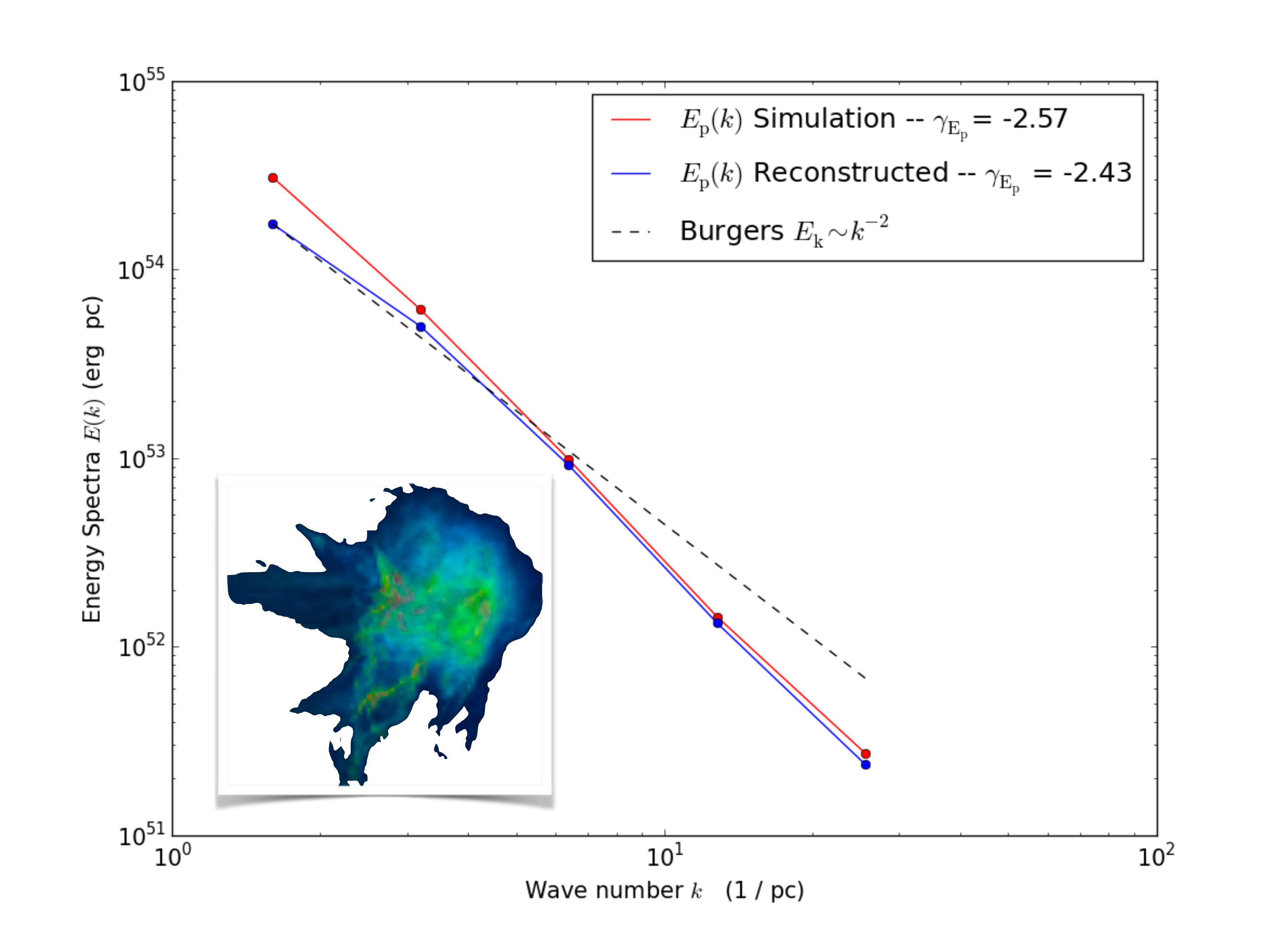}
\caption{\label{fig:comp}A comparison of the gravitational energy spectrum
computed from a Mach 5.6 compressible turbulence from
\citet{2010A&A...512A..81F} and the gravitational energy spectrum reconstructed
from the simulated observations. Scaling exponents for both cases are obtained by fitting 
straight lines in the log-log space. The { density
structure generated in the } simulations have $E_{\rm p}\sim k^{-2.57}$  and
 the density
structure generated from the simulated observations we obtain $E_{\rm
p}\sim k^{-2.43}$.  The black dashed line is the kinetic energy power spectrum of Burgers turbulence. 
A volume
rending of the simulation data is included as an inset plot for reference.}
\end{figure}

We compute the gravitational energy spectrum  of the density
structures in   this turbulent flow, the results are presented in Fig.
\ref{fig:comp}.
The gravitational energy spectrum of the simulation can be described by $E_{\rm} \sim k
^{-2.57}$, steeper than what is seen in our observations, but a similar to that
of the California molecular cloud \citep{2016arXiv160304342L}. For clouds like
California, turbulence can provide support against
gravity.
But for clouds like those seen in the Orion complex, the cloud structure differs
significantly from the structures seen in a typical turbulent box, suggesting
that gravity is playing a role that is more important that the
California.

We compare the gravitational
energy spectrum computed directly from the simulation with the spectrum
computed based on 3D reconstructions of the 2D projected density distributions. 
{ We assume that the observations can
reliably trace the density distributions, and the effect of the distribution of
dust temperatures, the distribution of gas-to-dust ratios and radiative transfer
effects are not modelled. } 
{  This allows us to access the effect of density reconstruction on
the gravitational energy spectrum.} The results from the simulation and from the
simulated observation exhibit a high degree of resemblance. The gravitational energy spectrum extracted from the simulation is $E_{\rm} \sim k
^{-2.57}$, and the result from the simulated
observation is $E_{\rm} \sim k ^{-2.43}$.

\begin{figure}
\includegraphics[width = 0.5 \textwidth]{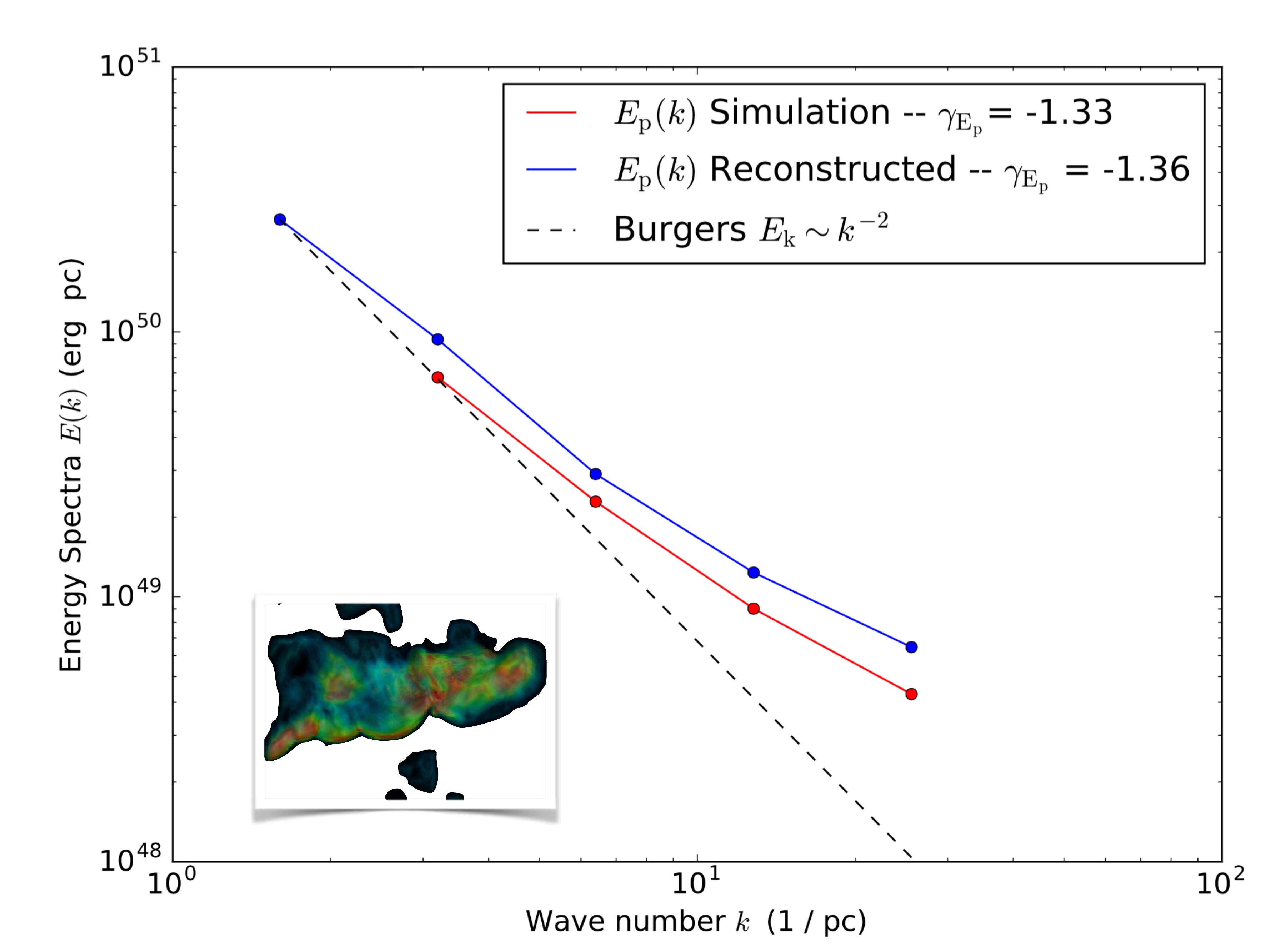}
\caption{\label{fig:comp:gray}{A comparison of the gravitational
energy spectrum computed from a $\sim\; \rm 1 pc$ large region obtained from
\citet{2015ApJ...807...67M} and the energy spectrum reconstructed from the
simulated observations. Scaling exponents for both cases are obtained by fitting straight lines in the log-log space. The { density
structure generated in the } simulations have $E_{\rm p}\sim k^{-1.33}$  and
 the density
structure generated from the simulated observations is $E_{\rm
p}\sim k^{-1.36}$.  The black dashed line is the kinetic energy power spectrum
of Burgers turbulence.
A volume
rending of the simulation data is included as an inset plot for reference.}}
\end{figure}

{
To test the performance of our method under a regime where gravity dominates, we
use the simulation from \citet{2015ApJ...807...67M}. The simulation includes
self-gravity, and we take a snapshot at 2 times the crossing time when gravity
dominates on small scales. We take a subregions of a size of $\sim 1\;\rm pc$,
and compare the gravitational energy spectrum derived directly from the simulations
with the one reconstructed from the simulation observation. The results are
presented in Fig. \ref{fig:comp:gray}. The gravitational energy spectrum of this
gravity-dominated simulation is $E_{\rm} \sim k ^{-1.33}$, which is shallower
than the energy spectrum of the Burgers turbulence. From the analysis of
gravitational energy spectrum alone, one would be able to conclude that gravity
is dominating the evolution of the system provided that the system is viralised
on the large scale.
In both the turbulence-dominated case and the gravity-dominated case, our
method performs to an accuracy of 5 \%,
and is able to constrain the variations of the gravitational energy spectrum to
a good accuracy.
}
 
%

  The reason for this accuracy of the reconstruction should be discussed
briefly. \cite*{2014Sci...344..183K} made a thorough study of the accuracy of
density reconstruction on the resulting volume density PDF. Although here we are
mainly concerned with the impacts of density reconstruction on the resulting
gravitational energy spectrum. Since the major difficulty for both cases is the
line-of-sight confusion, how the reconstruction depends on the underlying
parameters (e.g. the Mach number and the mode of forcing) should be similar.
\cite*{2014Sci...344..183K} found that the density reconstruction performs
better for close-to-solenoidal forcing. In this work, to test the robustness of
our density reconstruction, we use a simulation with fully compressible forcing.
It is also believed that \citep{2008ApJ...688L..79F} the interstellar medium 
is dominated by
compressible rather than solenoidal modes,  consistent with our choice.
According to our result, the density reconstruction allows one to evaluate
the gravitational energy spectrum to an accuracy of 5 \%. 
This should be taken as a fiducial  value, and might change with the complexity
of the underlying density structures.


The 5 \% accuracy of the estimated
gravitational energy spectrum can be understood
analytically. Suppose that we are measuring the gravitational energy spectrum of
a region from 1 pc to 0.1 pc. 
{ In our improved density reconstruction method, filamentary
structures are preserved. The aspect ratios can
be slightly underestimated, but this should not contribute much to the error.
This is because gravitational energy is not sensitive to the aspect ratio.} An
aspect ratio of 10 only changes the gravitational energy estimate by a factor of 1.4.
\citep{1992ApJ...395..140B,2016arXiv160304342L}. { For an individual
structure}, at most, our gravitational energy spectrum can deviate from the real
one by a factor of $\log(1.4) /  \log(10) \sim 15 \%$. This provides a safe estimate of the 
uncertainties of the estimated slopes of the gravitational energy spectra.
This is merely an upper limit.
Thus, we expect our method to be more accurate than this.
For a typical
molecular cloud, there exists a large number of such sub-regions.
Assuming these regions are randomly oriented,
the error of the estimate should decrease with increasing number $N$ of
the regions as $\sim 1/\sqrt{N}$. It is thus understandable that a $\sim 5 \%$
 accuracy has been
achieved.

Another observational constraint should be discussed: In our Orion data,
after generating the size map, a diffuse background with a column density of
$\sim 0.008\;\rm g\, cm^{-2}$ is not taken into account in the density
reconstruction process. This background results from a diffuse
 component that is larger than e.g. half the box size. It is difficult to tell
observationally if this background emission is from the cloud, 
or whether it is a contribution from foreground/background structures. Nevertheless, because we have limited
ourselves to scales that are smaller than half of the box size, our results are not sensitive
to this.

s

\begin{table}
\begin{tabular}{ l|l l l }
\hline
Cloud Name &  $\gamma_{\rm E_{\rm p}} $ analytical & $\gamma_{\rm E_{\rm
p}}$ this work 
\\
\hline
Orion A & 1.89 & 1.88 \\
\hdashline
 Orion B  & 2.0 & 2.17\\
\hline
\end{tabular}
\caption{ Scaling exponents $\gamma_{\rm E_{\rm p}}$ of the gravitational energy
spectra for clouds in the Orion complex ($E_{\rm p}(k) \sim k^{-\gamma_{\rm
E_{\rm p}}}$).
``$\gamma_{\rm E_{\rm p}} $ analytical'' stands for the  $\gamma_{\rm
E_{\rm p}} $ inferred from the surface density PDF, using the analytical
formulation derived in \citet{2016arXiv160304342L}, and ``$\gamma_{\rm E_{\rm
p}}$ this work'' shows the scaling exponent of the
gravitational energy spectra derived from the observations. \label{tbl:comp} }
\end{table}

\section{Comparison with other methods}\label{sec:comp}
Previously, methods have been proposed to evaluate the impacts of
gravity on cloud evolution. Methods such as the virial parameter allows
direct comparisons between gravitational energy and e.g. energy from
turbulent motion.
These methods only allow one to quantify gravity on a given scale, but not
across different scales. Because of the complexity of the star formation
process, to fully understand cloud evolution one needs multi-scale methods. Among these, the most commonly-used one
is the correlation function. In numerical simulations, it has been demonstrated that the density correlation function can be used to probe the
gravitational collapse of molecular clouds
\citep{2013ApJ...763...51F,2015ApJ...808...48B,2012ApJ...750...13C}. 
Because of gravity, the power spectrum of density can be very shallow and even
positive-valued in numerical simulations. This has been suggested by the authors
as a clear signature of gravitational collapse.

Indeed, the effect of gravity in cloud evolution can be observed both as
variations in the gravitational energy spectrum and as variations in the density
power spectrum. 
Both methods have their own advantages. Here, we emphasise what can be achieved
additionally with the gravitational energy spectrum:
First, the density power spectrum is a statistical measure, and for the results
to be meaningful, it is assumed that the statistics must be close to Gaussian (or log-normal), and
the background is approximately statistically homogeneous \footnote{These limitations are sometimes mentioned in the literature.
One can see e.g. \citet{2016ApJ...818..178L} where the authors propose to
combine complementary statistical
measures. The gravitational energy spectrum and the density power
spectrum does compliment each other in this sense.}.
{ Our method is not statistical. It is based on the wavelet
transform, and the energies are evaluated through integration over space
(Eq. \ref{eq:int})}.
In
the cases where the cloud is significantly inhomogeneous (e.g.
for filaments like the Musca molecular cloud), one can still apply the gravitational energy spectrum to study the
scale-dependence of the gravitational energy, but should not apply the
density power spectrum blindly. Correlation measures such as the  power spectrum
are also blind to the phase correlations  of the
independent Fourier modes \citep{2004MNRAS.350..989C}.
Second, the gravitational energy spectrum is a measure of gravitational energy as a function
of scale. The method is proposed with the intention of comparing various energy
terms in molecular clouds (e.g. turbulence, gravitational and magnetic, and
perhaps thermal energy), { and it measures the amount of
gravitational energy as a function of scale. Combined with complimentary
measures of velocity structures, we will be able to constrain the interplay
between turbulence and gravity to a better accuracy.} 

One should also note that, obviously, the gravitational energy spectrum and the
density power spectrum are different measures, and regions  that have the same
density power spectrum can have different gravitational energy spectra, and vise versa. A
proper combination of these two will provide a better description of the
structure of the ISM. This should be explored in the future.

Methods that constrain gravity by synthesising the position and the velocity
information include the Dendrogram method and the G-virial method. The
Dendrogram method \citep{2008ApJ...679.1338R} is a generic tool to describe
structures seen in observations. \citet{2009Natur.457...63G} applied the method
to molecular line mapping observations  in the PPV (positions-position-velocity)
space to constrain gravity on multiple scales, and found that gravity is universally important. The G-virial
\citep{2015A&A...578A..97L} method provides constraints on gravity by
generalising the concept of gravitational potential to 3D PPV space. The authors compute the
virial parameter of the clouds based on the G-virial maps from $\sim 0.1\;\rm
pc$ to a parsec.
They found that the clouds are gravitational bound on multiple scales.
It is interesting to note that a similar conclusion has been reached with two
almost independent methods. Compared to them, the gravitational energy spectrum
offers a different perspective, as it provides constraints of gravity in terms of energy. This would allow one to compare it
with turbulence energy and magnetic energy provided that similar
constraints are available. 

Recently, \citet{2016arXiv160305720L} proposed an acceleration mapping method to
constraint the effect of gravity in accelerating gas. The method computes the
acceleration induced by large-scale gravity using observational data. They found
that acceleration tends to be concentrated in localised regions --
as pointed out in \citet{2004ApJ...616..288B,2007ApJ...654..988H}. At the
current stage, this  method does not provide constraints on gravity in a multi-scaled fashion.

\section{Conclusions}\label{sec:conclu}
Gravity plays an important role in the evolution of the
molecular ISM.
In a previous work \citep{2016arXiv160304342L}, we proposed a measure called gravitational
energy spectrum to quantify the importance of gravity on multiple physical scales.
 In this
work, using a wavelet-based decomposition technique, we
derive gravitational energy spectrum of the Orion A and the Orion B molecular
cloud.
The derived energy  spectra cover the range from 0.1 pc to 10
pc, and provide constraints on the importance of gravity on these scales. 

It is found that gravitational energy spectra have power law-like shapes. At
sub-parsec scale, the Orion A molecular cloud has $E_{\rm p}(k)\sim k^{-1.88}$
and Orion B has $E_{\rm p}(k)\sim k^{-2.09}$. These
scaling exponents agree with our earlier analytical estimates
\citep{2016arXiv160304342L} to an accuracy of 10 \%.
The fact that
these scaling exponents are close to the exponents of the kinetic energy
power spectra of turbulence (where $E_{\rm p}\sim k^{-2}$) indicates a
multi-scale equi-partition between turbulence and gravitational energy. It also 
provides a clear
evidence that gravity is able to counteract effectively 
against turbulence from the cloud scale down to $\sim 0.1\;\rm pc$. 
For the Orion A molecular cloud, if the cloud as a whole is close to being
gravitationally bound, gravity inevitably dominates over turbulence inside the
cloud. 
 We also computed the gravitational energy spectrum from the
density structure generated from  a simulation with a Mach 5.6 compressible
turbulence, and found  $E_{\rm p}\sim k^{-2.43}$, which is significantly
steeper than clouds in the Orion complex but is comparable to e.g. the
California molecular cloud \citep{2016arXiv160304342L}.

{ 
Our work provides a multi-scaled view of molecular cloud dynamics. It
demonstrates that turbulent and gravitational energy reaches a rough
equi-partition, and proves that gravity is playing an important role in the
evolution of turbulent star-forming molecular clouds form the cloud scale down
to $\sim 0.1\;\rm pc$. This demands a better understanding of the interplay
between turbulence and gravity (perhaps along the path of
\citet{1995A&A...303..204V,1987A&A...172..293B})}. Perhaps the dynamics of
star-forming molecular clouds could better be described by the hierarchical gravitational collapse model \citep{1953ApJ...118..513H,1993ApJ...419L..29E,2003ApJ...585L.131V,2013ApJ...773...48B,2004ApJ...616..288B}.
Application of the technique to different numerical simulations will help to
clarify this issue, and are planned as our future work.

We note, however, that in general molecular clouds have diverse structures. 
Seen from the perspective of the gravitational energy spectrum, for clouds like
Orion A and Orion B (named g-type in \citet{2016arXiv160304342L}), gravity is
important,  for non-star-forming clouds like the California molecular cloud (named t-type in
\citet{2016arXiv160304342L}), turbulence can dominate. 
 
\appendix

\section*{Acknowledgements}
Guang-Xing Li is supported by the Deutsche Forschungsgemeinschaft (DFG) priority
program 1573 ISM- SPP. The paper also benefits from a careful review from the
referee.

\appendix
\section{Multi-scale decomposition of the surface density
map}\label{sec:size:map} \cite*{2014Sci...344..183K} proposed to decompose the
surface density maps into contributions from structures on different physical
scales using the \emph{a trous} wavelet transform, in order to construct the
so-called size map. This can be implemented as a series of convolution and subtraction
operations.
For casual purposes, this approach can provide reasonable results. However,
based on filtering in the Fourier space, wavelet transforms are known to
produce negative artefacts when the input data is not sufficiently smooth and regular. We first tried to decompose the
Herschel-Planck surface density map of Orion A, and found that these negative
artefacts contain a significant fraction ($\sim$ 20 \%) of the total mass. The articacts
are particular pronounced around structures where the column densities are
significantly enhanced (e.g. around the cores). We thus developed a 
new iterative algorithm to decompose the emission map into contributions from
structures on different physical scales.

We describe its
implementation briefly. The method works by recursively removes significant
structures at different resolutions. For each map, the resolution of interest starts from the
pixel size to about half of the map size. The coarser resolution is always twice
the finer resolution. The procedure starts with the finest resolution.

For a given resolution level $l = 2^k$ ($k=0,\, 1,\, 2,\, ,\ldots$), the
emission map $I(x, y)$ is smoothed with a Gaussian kernel whose size (measured in $\sigma$)
is 2 times of the current resolution, yielding $I_{\rm sm}(x, y)$. Then significant structures
are identified as those where $(I(x, y) - I_{\rm sm}(x, y))/ I_{\rm sm}(x, y)$
is larger than a given threshold. 
We 
experimented with different thresholds, and found that a threshold of $2$ gives
reasonable results. A change of the threshold does not change the results
significantly. These significant structures are stored as a separate array
($I_l(x, y)$, where $l$ represents the current resolution), and are subtracted from the map $I(x, y)$.
\emph{This process is repeated until all significant structures have been
removed from $I(x, y)$ and stored in the array $I_l(x, y)$.} Afterwards, the
subtracted emission map contains structures that are $\gtrsim 2$ times the
current resolution level. This residual map is smoothed with a Gaussian kernel
to ensure that it is sufficiently smooth at the next resolution level, 
 and is passed
to the next level of resolution where $l = 2^{k + 1}$.

\begin{figure*}
\includegraphics[width = 1 \textwidth]{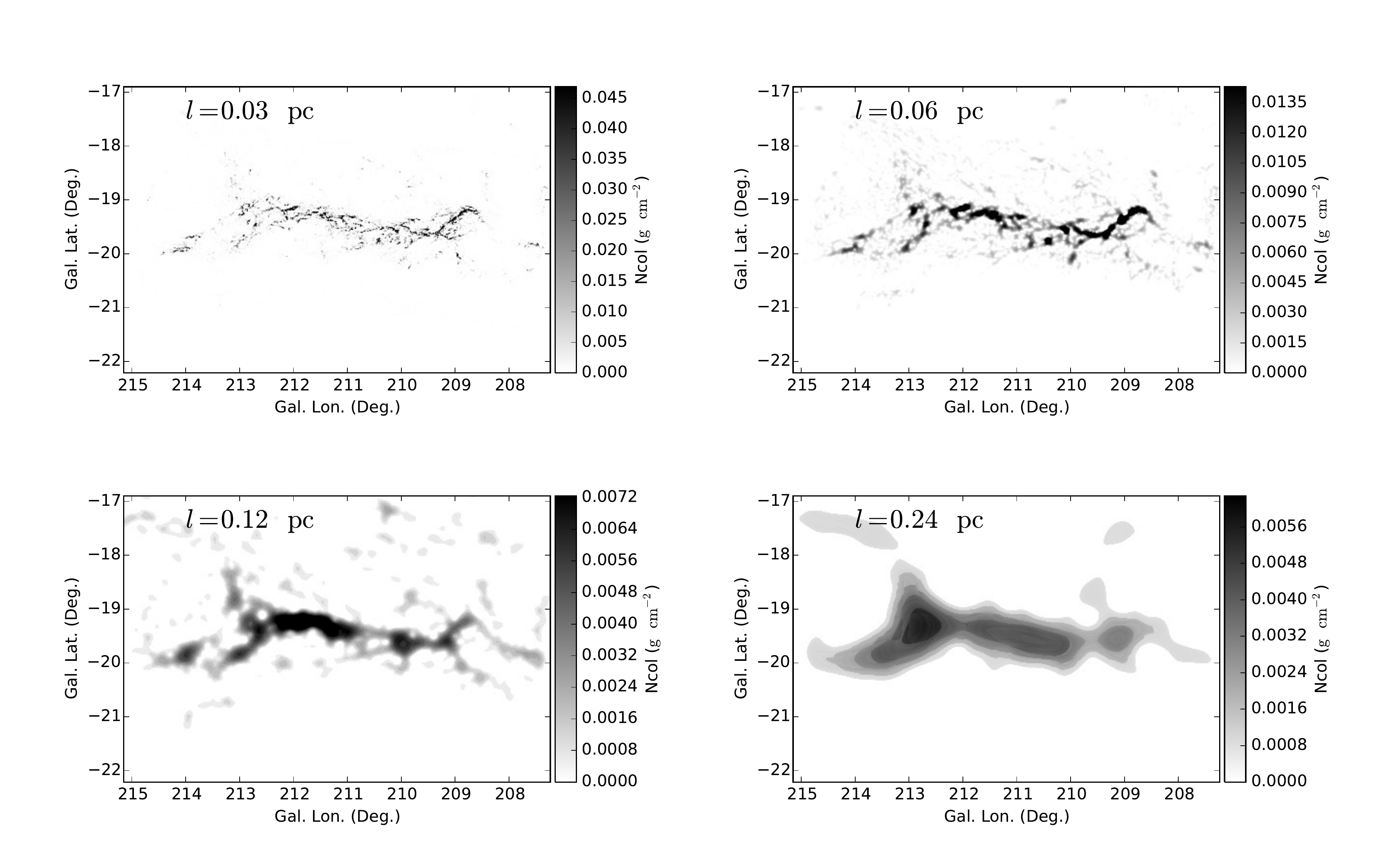}
\caption{ A few channels of the size map of the Orion A molecular cloud. 
Different maps contain structures of different physical sizes. The relevant
physical scales are indicated in the panels. Here $l$ is the dispersion of the
Gaussian kernel used for computing the size map. See Sec. \ref{sec:size:map} for
details.\label{fig:size_map}}
\end{figure*}

In Fig. \ref{fig:size_map} we present the size map of the Orion A molecular
cloud produced with our recursive algorithm. Different from the wavelet-based
decomposition theme, the new iterative algorithm does not produce any negative
artefacts.


\section{Contour-based volume modelling}\label{sec:3d}
In \cite*{2014Sci...344..183K}, {  the volume density modelling
consists of two steps: the first step is to decompose the map into different
structures, and the second step is to related the identified 2D structures
into 3D structrues.}

After decomposing the input density map into contributions
from different sizes, the clumpfind algorithm were used to originally identify
structures that are later used in reconstructing a 3D density distributions (this step was
originally named as ``volume density modelling'' by these authors). However, in
their original formalism, the structures are identified by the clumpfind algorithm
\citep{1994ApJ...428..693W}, and were treated as ellipsoids. This introduced
sharp edges to the resulting 3D volume density map.
{ The structures identified by ``Clumpfind'' algorithm blindly should
 also be considered
 to be correct only in a statistical sense \citep{2009ApJ...699L.134P}.}

For each surface density map $I_l(x, y)$ representing emission contributed from
structures of sizes $\sim l$, our method contour-decompose starts from maximum
surface density threshold, identifies the significant structures above this
threshold, registers them in a catalogue, removes these structures from the
emission, and then further decreases the threshold to include less significant
structures.
This continues until a minimum level is reached. In our calculations, the
minimum level above which structures are considered to be significant is set to
be 5 \% of the maximum level, and for each map we choose 50 equally 
spaced contour levels.

{  To put it simple, in our decomposition, the region is split
based on a set of contours where the gas inside each contour is taken as an independent structure. Our contour-decompose maximises the use of information in the
observed map, and allows one to reconstruct a 3D density map while preserving many of the
filamentary structures that also obvious in 2D. }

For each of the identified structure, its extent along the line of sight is
estimated by diagonalizing the tensors of the second moments of the pixel
positions. {  A 3D density distribution is reconstructed from all
the identified structures from our contour-decompose method.} 
%
The mass
distributions along the $z$ direction are modelled by assuming that they follow
Gaussian profiles where the dispersion $\sigma$ is equal to the estimated
extent along the third dimension.

For regions like the Orion A, around $10^4$ structures
are identified and folded into 3D to reconstruct the volume density map. The
reconstructed volume density map is in general very smooth.

\bibliography{paper}

\end{document}